# Astro2020 APC White Paper

# SmallSats for Astrophysics

**Paper Type:** Activities; **Thematic Area:** Electromagnetic Observations from Space

**Principal Author:**
David R. Ardila
Jet Propulsion Laboratory, California Institute of Technology
Email: David.r.ardila@jpl.nasa.gov; Phone: +1-626-658-0693

**Co-Authors:** Anthony Freeman (Jet Propulsion Laboratory), Todd Gaier (Jet Propulsion Laboratory), Varoujan Gorjian (Jet Propulsion Laboratory), Evgenya Shkolnik (School of Earth and Space Exploration, Arizona State University), Scott Wolk (Harvard-Smithsonian Center for Astrophysics)

**Endorsers:** Rachel Akeson (Caltech-IPAC); Amirnezam Amiri (Institute for Research in Fundamental Sciences); Daniel Apai (University of Arizona); Geert Barentsen (Bay Area Environmental Research Institute); Rachael Beaton (Princeton/Carnegie Observatories); Derek Buzasi (Florida Gulf Coast University); Douglas Caldwell (SETI Institute); Thayne Currie (NASA-Ames); William Danchi (NASA Goddard Space Flight Center); Thomas Greene (NASA Ames Research Center); Carey Lisse (Johns Hopkins University Applied Physics Lab); Joe Llama (Lowell Observatory); Patrick Lowrance (Caltech-IPAC); Michael Meyer (University of Michigan); Paulo Miles-Páez (University of Western Ontario); John Monnier (University of Michigan); Peter Plavchan (George Mason University); Sam Ragland (W.M. Keck Observatory); Paul Scowen (Arizona State University); Sara Seager (Massachusetts Institute of Technology); Michael Shao (JPL); J. Allyn Smith (Austin Peay State University); Erin Smith NASA (Goddard Space Flight Center); Arif Solmaz (Çağ University); Gerard van Belle Lowell Observatory); Michael Zemcov ((Rochester Institute of Technology)

**Table of Contents**

# Key Issue

The commercial SmallSat industry is booming and has developed numerous low-cost, capable satellite buses. Together with the large number of launch vehicles also being developed, SmallSats provide astronomers with new opportunities for accessing space. For some years now, NASA Astrophysics has solicited SmallSat missions through the Mission of Opportunity (MoO) call and the Astrophysics Research and Analysis (APRA) Announcements of Opportunity (AO), but this has resulted in only one launched SmallSat to date.

SmallSats can be used as vehicles for technology development or to host science missions. Missions hosted on SmallSats can answer specific science questions that are difficult or impossible to answer with larger facilities, can be developed relatively quickly, serve to train engineering and scientists, and provide access to space for small institutions. SmallSats complement larger Astrophysics missions and allow the broader community to test new ideas at the bottom of the market, creating new capabilities which find their way to larger missions.

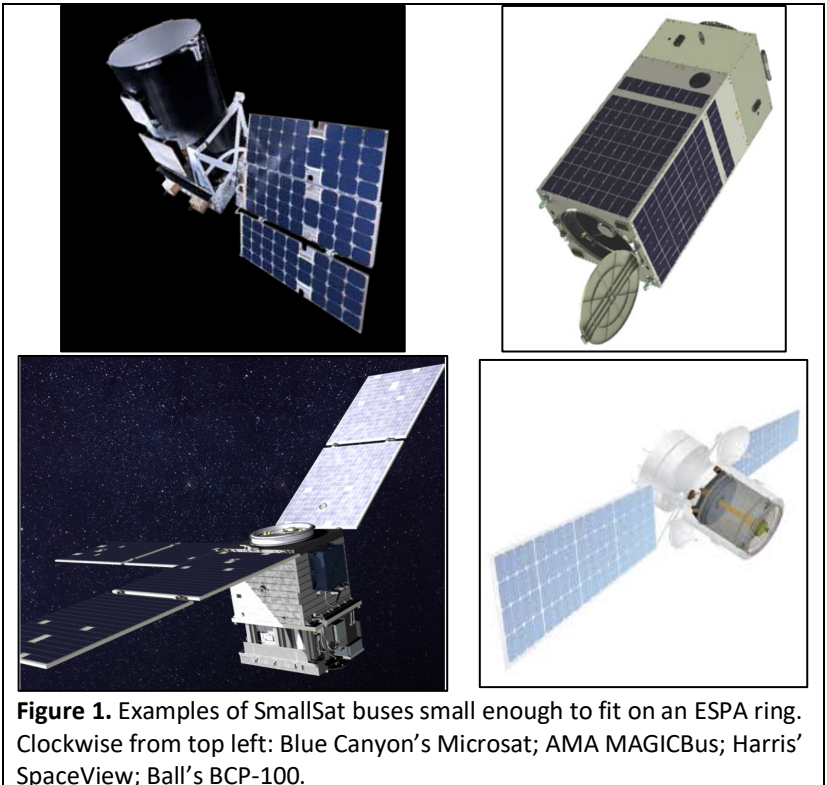

**Figure 1.** Examples of SmallSat buses small enough to fit on an ESPA ring. Clockwise from top left: Blue Canyon's Microsat; AMA MAGICBus; Harris' SpaceView; Ball's BCP-100.

However, without flight opportunities to mature technologies, missions hosted on SmallSats are likely to be considered high risk, and face long odds being selected for implementation. Currently, NASA Astrophysics does not provide flight opportunities that would allow technology maturation of instrument systems or concepts of operations.

Our primary suggestion is that NASA decouples science and technology for SmallSats by creating a technology-based SmallSat AO, modeled after the Earth Sciences InVEST call. Such AO would help reduce the new technology risk for science missions of any size.

We also suggest that NASA provides additional science-driven SmallSat opportunities at the ~$12M funding level, provides access to new launchers free of charge to proposers, and re-structures the solicitation AOs so that SmallSats do not compete with other mission classes such as balloons.

# Introduction

The small satellite (SmallSat) industry is booming, with over 1,000 SmallSats launched in the last 5 years, and a forecast of $30B dollars in business predicted for the upcoming decade (Sahdeva 2019, Adamowski 2017). NASA itself spends ~$100M/year in SmallSats (Zurbuchen, 2018). The large number of satellite providers has resulted in a variety of low cost, reliable, and flexible buses (Figure 1).

This is complemented by new opportunities to launch SmallSats as primary payloads to custom orbits, facilitated by the large number of small rocket launchers in development (Ponnappan 2018). SmallSats can also be launched as secondary payloads to variety of orbits.

This supply is matched by strong demand, as access to space remains crucial for astronomy. Except for particular windows in radio, optical, and near-infrared spectral ranges, Earth's atmosphere is opaque or highly variable. Even within the optical window, high photometric precision, and high spatial resolution imaging over wide areas is made very challenging by atmospheric seeing. The day-night cycle of the Earth makes continuous time-domain (TD) observations difficult, requiring facilities all over the globe, with instruments that need very precise relative calibration.

**Table 1. NASA Astrophysics spacecraft launch opportunities.** The specific language regarding SmallSats changes in every AO. Here we quote from the most recent available AO.

| Type | Development Time | Cost | Examples |
|---|---|---|---|
| Flagships | ~Decades | ≥$5B | Hubble, JWST, Chandra, etc. |
| Probes (1) | Likely ≥10 yrs | ~$1B | None |
| Medium Explorer (MidEx) | ~8 yrs. | $250M (FY2017) (2) | FUSE, WMAP, WISE, TESS, etc. |
| Small Explorer (SmEx) | ~5-8 yrs. | $145M (FY2020) (2) | GALEX, NuStar, IXPE, etc. |
| Small Complete Mission (SCMs) | ~5 yrs. | $75M (FY2020) (3) | None |
| "SmallSats" (includes CubeSats ≤12 U) | ~5 yrs. | $35M (FY2020) (4) | None |
| Possible new class (5) | | $12M | None |
| CubeSats ≤12 U (APRA) | ~3-5 yrs. | ~$5M (4) | 1 launched, 4 in development |

Notes:
(1) Proposed class
(2) Launch vehicle costs outside the cap
(3) Proposer must pay for launch, unless hosted on ISS or goes to lunar gateway
(4) Free launch as secondary payload
(5) The funding gap between "SmallSats" and APRA CubeSats suggests a new possible class at the ~$12M level

In addition to their utility as science platforms, or as technology maturation vehicles, SmallSats provide a training ground for the next generation of engineers and scientists. The fast development times allow them to gain valuable experience by participating in the full life cycle of a mission. Frequent small missions yield benefits to the larger projects through streamlining and bolstering of flight practices.

SmallSats facilitate the democratization of space, allowing for broader access to space-based science and technology. They add a diversity of voices and perspectives beyond the large centers and universities that dominate the narratives surrounding space astrophysics.

NASA provides astronomers with a variety of opportunities for access to space (Table 1). When it comes to SmallSats, the MoO element of the Explorer AO has solicited them (as 'nanosats', 'microsats', or CubeSats) since 2007. APRA has solicited CubeSat investigations since 2012.

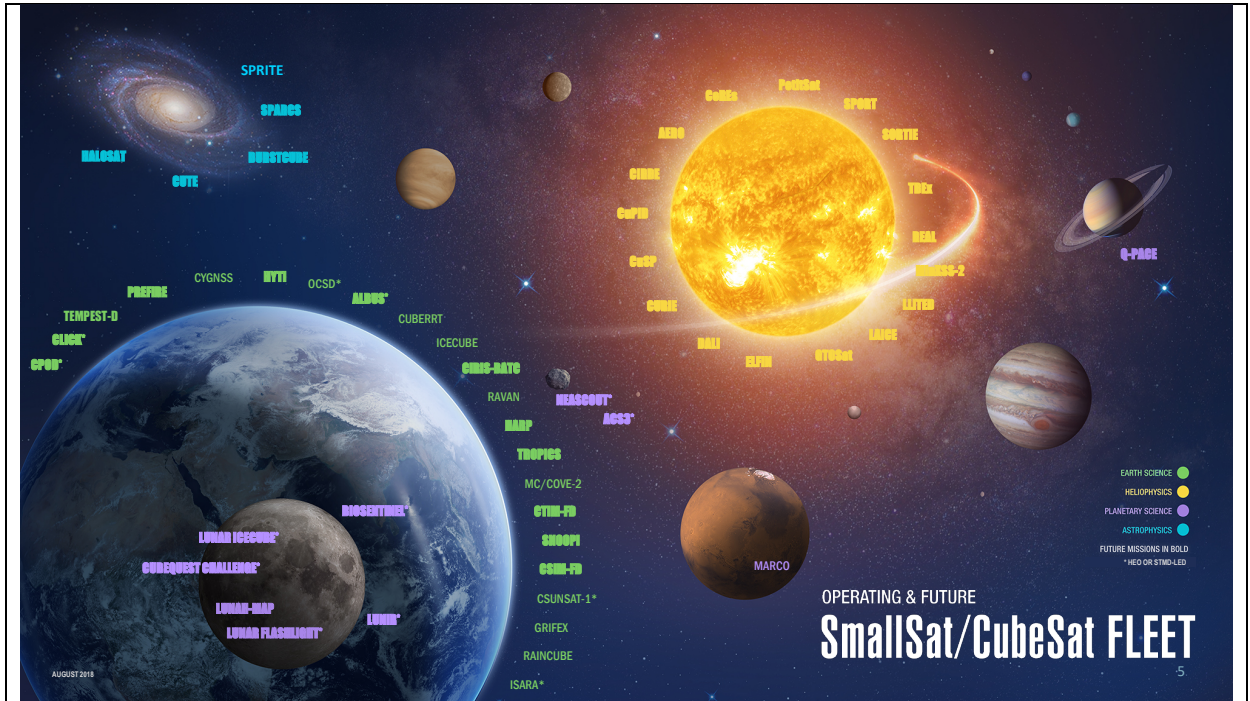

**Figure 2**. NASA SmallSat fleet. All the Astrophysics SmallSats are 6U CubeSats, and only one (Halosat) has launched. From T. Zurbuchen, NASA Science -SmallSat Strategy, SmallSat Conference (2018). Graphic modified to add SPRITE.

And yet, in spite of the availability of satellite buses, the needs of the community, and the opportunities for access to space, NASA Astrophysics has so far launched only one SmallSat, the 6U CubeSat HaloSat (Kaaret 2019), although four additional CubeSats are currently in development (Figure 2). As a community, Astrophysics has not taken full advantage of the opportunities offered by the developments in the commercial sector.

One of the main reasons behind the paucity of SmallSats in Astrophysics is the lack of risk reduction opportunities, and we discuss that here. Other potential reasons, briefly addressed here, include incorrect costing and uneven burdens within the program.[1] Therefore, while SmallSats have the potential to revolutionize Astrophysics, they remain mostly untried due to the way NASA Astrophysics manages and funds access to space.

---

[1] For example, on 2016 SmallSats competed under the same Mission of Opportunity AO with balloon investigations, and sounding rockets. However, balloons and sounding rockets have much shorter lifetimes, their "buses" are lower cost, and risk profiles are fundamentally different than SmallSats'.

## What is a Small Satellite?

We adopt the definition of a SmallSat as having a total mass ≤180 kg[2]. This used to be the mass limit for spacecrafts to be launched as secondary payloads (attached to an ESPA ring[3]) although larger masses are now possible. The 2019 Astrophysics MoO Announcement adopts the more restrictive definition of a SmallSat as a mission whose life cycle cost is less than $35M[4].

In general, an astrophysics mission with 180 kg satellite is likely to cost more than $35M. For comparison, the Galaxy Evolution Explorer (GALEX), with a telescope diameter of 50 cm, had a total mass of 277 kg, and cost ~$170M (FY2020, Wall 2013).

At the upper mass end, a SmallSat would have dimensions 61 cm x 71 cm x 97 cm, which could fit a 60 cm diameter-class, single aperture, telescope. At the lower end, the Starshot Initiative recently launched six 4 grams satellites (Crane 2017). The most common SmallSats are CubeSats, sized in units of 10 cm x 10 cm x 10 cm, each 1U has a mass of ~1.3 kg.

The cost of SmallSat buses depends on their capabilities but querying to multiple vendors indicates that one could buy a high-performing SmallSat bus for as low a ~$5M, not including Integration and Testing. For a $35M cost cap, this is 14% of the total cost. For comparison, a satellite bus to fit a Small Explorer payload is ~$45M, which is 30% of the cost[5]. In other words, a SmallSat mission can spend relatively more in science and payload than a Small Explorer, everything else being equal.

It is worth pointing out that given the paucity of launched missions (one), it is unclear how much an Astrophysics SmallSat designed to fulfill science requirements would actually cost. For a wide variety of mission classes, a commonly used rule of thumb is that the price of spacecraft plus payload should comprise half of the total mission cost, not including the launch vehicle. Therefore, a $35M cost cap should afford $18M worth of payload and spacecraft.

Table 1 shows that, except for APRA, NASA uses a factor of ~2 scaling between the different mission classes. The table shows a gap between the $35M "SmallSats" and the APRA CubeSats. This provides a potential opportunity for a new science-driven class of SmallSats (probably CubeSats).

## SmallSats Technology Gaps

There are areas in which the needs of astrophysics push against engineering capabilities of SmallSats. Table 2 provides a list of general technologies of broad applicability to astrophysics. This is not an exhaustive list, but it captures some of the current concerns in the community. As the CubeSats currently on the pipeline and the AS3 responses show (see section *Science and Risk*), these gaps have not stopped the community from developing new concepts.

---

[2] https://www.nasa.gov/content/what-are-smallsats-and-cubesats.
[3] Evolved Expendable Launch Vehicle (EELV) Secondary Payload Adapter (ESPA).
[4] Life cycle cost includes all costs, from formulation to close-out, including science analysis.
[5] FY2020, based on vendor quotes.

| Table 2: SmallSat Technology Gaps for Astrophysics | | | | |
|---|---|---|---|---|
| **Technology** | **Metric** | **Observations Enabled** | **State of the Art** | **Comment** |
| **Deployable Apertures** | Effective aperture size | All | 50 cm Ka antenna (6U RainCube 2018). | No UV-Optical-IR deployable aperture demonstrated. |
| **Compact active cooling systems** | System/subsystem temperature | Infrared observations | None | CIRaS to use a Ricor K508N cryocooler; BIRCHES to use AIM SX030 |
| **Wavefront Control** | Wavefront Error | High-contrast imaging | None | Deformable mirror with 140 actuators to be demonstrated by 6U DeMi (2019) |
| **Optical Downlink** | Downlink rate | Large area imaging | 100 Mbps (Aerocube 7B, 7C) | TBIRD concept to demonstrate 200 Gbps (2019). |
| **Specialized Orbits** | Orbit parameters | Time-domain; Low background (X-ray, UV, IR) | LEO | Limited by the use of secondary payload slots. |
| **Constellations** | Number of spacecraft | Follow-up of episodic events; RF observatories | None | SunRISE: Constellation of 6 CubeSats; Concepts for larger HF constellations in lunar orbit. |
| **Jitter control** | High frequency pointing error during integration times | High resolution imaging; Spectroscopy; astrometry | 1σ=0.5" (ASTERIA, with active jitter control) | Jitter control with start trackers has been demonstrated by MINXSS (1σ =5"-15") |

For photon-starved applications, fitting a large aperture in a small bus requires consideration of packing and optical prescription: a fast telescope feeding a compact instrument can fit on a small package but it will not be ideal for high resolution imaging. The largest aperture currently planned for a SmallSat is CUTE's square mirror at ~20 cm x10 cm. Concepts for deployable apertures (Champagne et al. 2014) exist but they remain at low Technology Readiness Level (TRL).

Low detector temperatures are necessary to reduce dark current values and increase charge transfer efficiency. For thermal infrared applications, low payload temperatures are needed to reduce background contribution. Compact, affordable active cooling systems remain untested in flight, though this may change in the next couple of years as tactical coolers are demonstrated on Lunar IceCube and HyTI, both carrying cooled Thermal IR imagers.

Pointing control provided by the spacecraft is already good enough for some applications. Star trackers in CubeSats have demonstrated long-term jitter control between 1σ =5"-15" (Mason et al. 2017). For comparison, GALEX's PSF is 1σ ~2"[6], close to CubeSat numbers. This is tight enough for photometric measurements of targets in well-known, relatively sparse fields. Slit spectroscopy or photometry of crowded fields requires better spacecraft pointing, or pointing management to be performed by the payload. For bright targets, pointing jitter as small as 1σ=0.5" has been demonstrated by the ASTERIA CubeSat, using active focal plane control (Smith et al. 2018).

---
[6] https://archive.stsci.edu/missions-and-data/galex-1/

The lack of specialized orbits is an insidious problem for SmallSats. For secondary payloads, the ISS orbit is the most likely destination. As is familiar to users of the Hubble Space Telescope, a spacecraft in low-Earth orbit experiences day-night cycles with a ~90 min period, thermal shocks, and target eclipses every ~45 min, in addition to a high Lyman-α and infrared backgrounds. This limits the photometric stability and precision, and imposes hard constraints on TD observations, one of the "killer apps" for SmallSats.

The 2019 MoO AO allows for a Geostationary Transit Orbit launches if available, launches to the Lunar Gateway, when developed, to the Sun-Earth L1 with IMAP, or L2 with PLATO. This is a larger landscape than previous AOs, but it still constrains SmallSat missions to a limited choice of orbits, determined by the needs and timing of other missions.

Launches to other, more scientifically useful orbits, such a Sun-Synchronous terminator orbit are rare. There are no Astrophysics SmallSats on the upcoming EM-1 mission, the maiden voyage of NASA's SLS rocket, though that affords an opportunity to escape the Earth's IR and UV background and provides better environment for TD observations. Earth-trailing (like Spitzer's and Kepler's) orbits, or high eccentricity (TESS', Chandra's) orbits are not available for secondary payloads. Orbits that would provide low particle background for X-ray observations (another killer app for SmallSats) have semimajor axes larger than the geostationary value, and are not available for SmallSats under any AO.

## Science and Risk

As Harwit (1984) argues, astronomical discoveries result from exploring new regions of the observational parameter space. TD observations, spectral ranges that cannot be reached from the ground, observations with specialized instruments, unique concepts of operations, and detailed observations of a particular sample of objects, provide the landscape of opportunities for SmallSats (Ardila et al. 2017, Shkolnik 2018).

The astronomical community has already concluded that SmallSats would be scientifically useful. On September 2017 NASA issued a Request For Information (RFI) on possible astrophysics projects at costs between $10M and $35M, and received 55 replies, on a wide variety of science topics. In April 2018, NASA released a Request for Proposals for Astrophysics Science SmallSat Studies (AS3), with a cost-cap of $35M. Thirty-five proposals were received and 9 were selected to receive study funds during 6 months (Loff 2018).

The selected proposals run the gamut from exoplanet to cosmology, demonstrating that in the eyes of the community, SmallSats can be used for all kinds of science and at all wavelengths. They include a starshade concept for high contrast imaging (mDOT), exoplanet finding via astrometry (MASS), exoplanet characterization via X-ray transits (SEEJ), X-ray detection of diffuse background (XQSat), radio observations of the 21 cm line (DAPPER), TD observations of the UV sky (GUCI++), and the X-ray sky (HSP), IR observations of galaxy clusters (ISCEA), and flying multiple X-ray spacecraft to improve spatial resolution (VTXO).

A detailed feasibility evaluation may conclude that the projects detailed in these proposals are risky, as they involve new buses, with new instruments and new observing modes. In the accepted AS3 proposals there is emphasis on techniques that have not been fully exploited from space (astrometry, polarimetry) or observation modes that have never been performed (distributed X-ray telescopes, starshade observations).

This is a problem because when it comes to access to space the current NASA process has very little appetite for risk. The feasibility of an investigation described in a proposal is judged in terms of technical, managerial and cost risk, in addition to the science merit and its implementation. If the mission is deemed to be Medium ("Mission design may be complex and resources tight.") or High Risk ("One or more problems […] of sufficient magnitude and complexity as to be deemed unsolvable [...]"), it will not be funded.[7]

## Can NASA Astrophysics build SmallSats?
As Figure 2 shows, NASA's SmallSat fleet includes a large number of Earth Science and Heliophysics missions. A significant fraction of the Earth Science missions were developed within the In-Space Validation of Earth Science Technologies (InVEST) program. InVEST uses CubeSats to advance instrument technologies that will provide measurements described in the Earth Sciences Decadal Survey, to advance the science goals described therein. The InVEST call has resulted in concepts long thought to be impossible with CubeSats: active radar, far infrared instruments, high resolution spectrographs, deployable apertures, etc.

Astrophysics does not have an InVEST equivalent that would provide a programmatic opportunity to flight a mission to validate subsystems, small instruments, and measurement techniques and concepts. The closest program within astrophysics is APRA, but even there CubeSat missions are expected to deliver both competitive science and technology advancement.

We believe that the lack of Astrophysics SmallSats is partially due to the lack of mature enough concepts that can compete in the NASA proposal process. SmallSat concepts need to prove that they are high-heritage and low-risk, in addition to show that their science is important and implementable. This is in conflict with using spacecraft that have not been previously used for astrophysics applications, proposing original instruments, or new measurement concepts.

Proposals become low risk by using previously tried buses, with instruments that depart little from what has been done before, performing similar measurements. When this is not possible, the proposers invest proposal funds in managerial oversight of critical items, invest in technology maturation, or downgrade the science investigation. These risk mitigation activities result in larger costs, longer development times, and less competitive science. By design, this is a very conservative process, which discourages innovation.

This conservatism is not necessarily a bad thing: we are all responsible for the judicious expenditure of taxpayer dollars. However, in order to reduce risk to acceptable levels, the proposal process should provide a pipeline to mature and test technology in space,. This pipeline is not currently available.  Flight opportunities that allow for technology maturation and proof-of-concept are crucial for the development of SmallSats in astrophysics.

---

[7] The quotes presented are taken from the 2019 Explorer AO, but similar criteria are mentioned in previous ones.

## Strategies

Overall, NASA Astrophysics SMD should be commended for their efforts to provide the community with SmallSat opportunities. Inspecting old AOs one can see an intent to make the calls responsive to changing opportunities. The Astrophysics SMD fosters an environment of open dialog, crucial to advance our common interests, and it is in that spirit that we offer the following suggestions:

- The Astro2020 Decadal survey should recommend that NASA provides flight opportunities that would allow to advance the maturity of astrophysics instruments, spacecraft systems/subsystems, or concepts of operations, for infusion into other missions, analogous to the InVEST AO. The technology gap list presented in Table 2 provides a list of those elements that need to be advanced first.
- NASA Astrophysics should take advantage of the development of new small launchers to manifest SmallSats as primary payloads and fully exploit the availability of launches that allow secondary payloads. NASA should provide launch opportunities to use these vehicles at no cost to the proposers, certifying launchers to use in possible SmallSat missions, or allow proposers to use commercially brokered rideshares.
- NASA Astrophysics should explore the feasibility of SmallSat missions at the $12M cost point, the logical missing step in the "access to space" ladder (Table 1). These are likely to be ambitious CubeSats that would deliver competitive science.
- NASA Astrophysics should consider re-structuring the AOs that request SmallSats so they compete only with other SmallSats. There are indications that Astrophysics Program Officers are already thinking about this, as the latest MoO AO does not solicit balloons or sounding rockets.